\documentclass[twocolumn,preprintnumbers,amssymb,aps,prd]{revtex4}

\newcommand{\bi}{\bibitem}
\newcommand{\be}{\begin{eqnarray}}
\newcommand{\ee}{\end{eqnarray}}

\begin{document}

\title{Constraints on temporal variation of fundamental constants from GRBs}

\author{C.~Bambi$^{\rm 1}$}
\author{A.~Drago$^{\rm 2,3}$}
\affiliation{$^{\rm 1}$Department of Physics and Astronomy, 
Wayne State University, Detroit, MI 48201, USA\\
$^{\rm 2}$Istituto Nazionale di Fisica Nucleare, 
Sezione di Ferrara, I-44100 Ferrara, Italy\\
$^{\rm 3}$Dipartimento di Fisica, 
Universit\`a degli Studi di Ferrara, I-44100 Ferrara, Italy}

\date{\today}

\preprint{WSU-HEP-0710}

\begin{abstract}
The formation of a strange or hybrid star from a neutron 
star progenitor is believed to occur when the central 
stellar density exceeds a critical value. If the transition
from hadron to quark matter is of first order, the event has 
to release a huge amount of energy in a very short time, and
we would be able to observe the phenomenon even if it is 
at cosmological distance far from us. Most likely, such violent 
quark deconfinement would be associated with at least 
a fraction of the observed gamma ray bursts. If we allow 
for temporal variations of fundamental constants like 
$\Lambda_{QCD}$ or $G_N$, we can expect that neutron
stars with an initial central density just below the 
critical value can enter into the region where strange or hybrid stars
are the true ground state. From the observed rate of long gamma
ray bursts, we are able to deduce the constraint 
$\dot{G}_N/G_N \lesssim 10^{-17} \; {\rm yr^{-1}}$,
which is about 5 orders of magnitude more stringent than 
the strongest previous bounds on a possible increasing $G_N$. 
\end{abstract}

\maketitle

{\sc Introduction --}
Search for temporal variation of ``fundamental constants''
has attracted a lot of interest in last years~\cite{uzan}.
Such a phenomenon is quite a general prediction of many
different frameworks (superstring theories, scalar-tensor 
theories, models in extra dimensions, etc.) when the dynamics 
of the whole universe is taken into account and a clear sign 
of violation of the Strong Equivalence Principle (SEP)~\cite{will}. 
On the other hand, the Einstein Equivalence Principle (EEP) 
-- which is the essence of the geometrical theory of gravitation 
-- refers only to non-gravitational physics and thus allows 
the Newton constant $G_N$ to be time dependent, but still 
forbids any temporal variation of non-gravitational parameters 
($\alpha$, $\Lambda_{QCD}$, $G_F$, etc.)~\cite{will}.

Since quantitative predictions are impossible at present, the 
common approach is based on model independent investigations 
of possible temporal evolution of physical quantities that 
should instead be constant in the standard theory and 
constraints are often reported assuming variations that are
linear in the cosmological time: indeed one can expect to
expand any false constant in a power series of time and 
that the linear term is the leading-order correction.
For this reason, essentially all the works in the literature
on this subject do not specify any theoretical framework
and the discussion is at a pure phenomenological level.
However, this approach cannot be always justified and hence, 
in general, one cannot directly compare bounds obtained 
today in laboratory with others coming from arguments involving 
the physics of the early universe.

The topic of time varying ``fundamental constants'' has 
received additional interest after the claims of some evidence of 
time variation of the fine structure constant $\alpha$~\cite{alpha} 
and of the proton to electron mass ratio $m_p/m_e$~\cite{pe-mass}.
There are indeed good theoretical arguments to believe that
a possible temporal evolution of $\alpha$ implies time shift
for other quantities as well~\cite{gut}. 
However the observational situation is quite controversial, 
since other groups have not confirmed the results~\cite{alpha2}.

A relevant quantity which sets strong interactions and 
hadron masses is the QCD scale $\Lambda_{QCD}$. The most
stringent constraint on its possible temporal variation 
comes from an analysis of isotopic abundances of the Oklo 
uranium mine in Gabon, a natural fission reactor that 
operated about $2 \cdot 10^9$ years ago for about $10^5$ years. 
Assuming a linear time dependence, one gets~\cite{oklo}
\be\label{bound-oklo}
\left| \dot{\Lambda}_{QCD}/\Lambda_{QCD} \right| 
\lesssim 10^{-17} \, {\rm yr}^{-1} \, .
\ee
Other constraints on $\dot{\Lambda}_{QCD}$ are usually
quite model dependent and, in any case, at least 2 orders
of magnitude weaker than the bound in~(\ref{bound-oklo}).

However, among all the ``fundamental constants'' a privileged 
role is played by the gravitational constant $G_N$, whose
variation violates only the SEP. 
The tightest constraint in the literature on a linear temporal 
evolution of $G_N$ is provided by the Lunar Laser Ranging 
experiment, which has monitored Earth--Moon distance for about 
30 years. The result is~\cite{llr}
\be\label{bound-llr}
\dot{G}_N/G_N = (4 \pm 9) \cdot 10^{-13} 
\, {\rm yr}^{-1} \, .
\ee
On the other hand, the earliest event in the history of the
universe which has left a reliable record of the past value 
of $G_N$ is the Big Bang Nucleosynthesis: from
the primordial abundance of $^4$He we can conclude that, when 
the universe was only 1 s old, i.e. $\sim 13$ Gyr ago, $G_N$ 
could not differ more than about 10\% from its present 
value~\cite{bbn}. Other interesting constraints on $\dot{G}_N/G_N$, 
at the level of $10^{-12} \, {\rm yr^{-1}}$ in the case of 
linear time evolution, can be deduced for instance from the orbital 
period of binary pulsars~\cite{pulsar}, from the comparison of 
masses of old and young neutron stars~\cite{thorsett} and 
from the so called gravito-chemical heating in neutron 
stars~\cite{gc-heating}.

In this letter, we consider the effects of a temporal 
variation of $\Lambda_{QCD}$ and $G_N$ on the equilibrium 
configuration of compact stars: indeed, non-zero $\dot{\Lambda}_{QCD}$ 
or $\dot{G}_N$ could change the true ground state of some
stars and induce conversion of neutron stars into strange 
or hybrid ones. Here the keypoint is that the transformation
probably involves a first order phase transition, leading to 
the release of a large amount of energy in a short time. 
This makes even small variation of $\Lambda_{QCD}$ or $G_N$ to 
produce spectacular events, observable at cosmological distances.
Taking into account the possible events which could be
somehow associated with these phenomena, we can estimate an
upper limit for the decreasing rate of $\Lambda_{QCD}$ or 
for the increasing rate of $G_N$.

To avoid any possible misunderstanding, a comment is in order 
here; we can only constrain dimensionless quantities, such as 
$\alpha$. When we talk about temporal evolution of dimensionful 
constants, like $\Lambda_{QCD}$ and $G_N$, we implicitly assume 
a particular system of units and again we constrain the 
dimensionless ratio to standards chosen as units and constant 
by definition~\cite{uzan}. This is particularly evident for 
example in some scalar-tensor theories of gravity, where one 
can work in the Einstein frame (where $G_N$ is constant) or in 
the Jordan frame (where particle masses are constant and $G_N$ 
can change), see e.g. Ref.~\cite{bellido}. However, this is
a more general result and one can always choose an arbitrary 
dimensionful quantity as standard unit and then compare it
with other quantities. And this is basically also what we do here: 
when we assume $G_N$ constant, we allow for variation of 
$\Lambda_{QCD}$, and when we assume $\Lambda_{QCD}$ constant we 
discuss the possibility of non-zero $\dot{G}_N$, but we compare 
always the QCD interaction with the gravitational one. For this
reason, the possibility of a compensation between the temporal 
variation of the QCD energy scale $\Lambda_{QCD}$ and of the
gravitational energy scale $M_{Pl} = G_N^{-1/2}$ has no physical
meaning: only their dimensionless ratio is an observable.
Of course, here we cannot exclude the possibility that both 
$\Lambda_{QCD}$ and $M_{Pl}$ change exactly in the same way 
with respect to another dimensionful constant, for example the 
vacuum expectation value of the Higgs field $v$, but such a
case sounds quite {\it ad hoc} and would require an unwanted
fine-tuning.

{\sc Compact stars --}
Compact stars are the end-product of heavy stars after
supernova explosion~\cite{sa-teu, glende}. Their mass is typically 
$\sim 1.5 \, M_\odot$ and their radius is about 10 km, so that 
they are extremely dense objects: in particular we expect that 
the central baryon density, $\rho_c$, is at the level of 2 -- 10 
times normal nuclear matter density $\rho_0 = 0.16 \, {\rm fm}^{-3}$.
At such high densities, matter would be made of neutrons, with 
a smaller fraction of protons and electrons. However, in 
the upper region of this interval, the ground state is 
probably represented by strange matter of $u$, $d$ and $s$ 
deconfined quarks: the latter are fermions and introducing a 
third flavor there are new low energy available states which 
can reduce the total energy of the system. If so, compact 
stars with a central density higher than a critical value 
$\rho_{crit}$ would be strange stars (stars made entirely of strange
matter) or hybrid stars (stars with a strange matter core
and an outer part of hadron matter), depending on QCD physics
and/or their mass. For a review, see e.g. Ref.~\cite{glende}.
Thanks to these extreme conditions, impossible to reach in
laboratories on the Earth, compact stars can really 
be seen as a unique opportunity
to investigate new physics~\cite{new-physics1, new-physics2}.

The results of this letter are essentially based on the 
following two assumptions: 
\begin{enumerate}
\item Strange or hybrid stars exist, but not all
the compact stars are strange or hybrid.
\item The transition from hadron to strange matter is
of first order.
\end{enumerate}
The first statement means that we assume that the population of 
compact stars is made of both neutron and strange/hybrid
stars. In other words, the critical density $\rho_{crit}$
has to be in the range of the typical central density of
compact stars, so that the ones with $\rho_c < \rho_{crit}$ 
are neutron stars, while if $\rho_c > \rho_{crit}$ they are 
strange or hybrid. As for the second hypothesis, it is 
indeed what is commonly believed, i.e. at high baryon 
density the phase transition from hadron to strange matter
seems to be of first order~\cite{phase-t}. Unfortunately, up
to now no model independent argument supporting that hypothesis
has been provided.

Let us now introduce the information which is needed to
constrain the temporal variation of the QCD scale and
of the gravitational constant. From considerations on
stellar evolution, we can say that about 0.1\% of the stars
in galaxies are compact ones, that is neutron or strange
stars (see e.g. Ref.~\cite{sa-teu}). This implies that there
are about $10^{8}$ compact stars in a typical galaxy
and that the whole visible universe contains a total number of
$N_{tot} \sim 10^{20}$~\footnote{Indeed, the number of stars in 
a general galaxy can be estimated to be about $10^{11}$,
while the number of galaxies in the visible universe is roughly 
$10^{12}$.}. As for their central baryon density 
$\rho_c$, we assume for simplicity that compact stars are 
uniformly distributed with $\rho_c$ in the range 
$(2 \, \rho_0, \, 10 \, \rho_0)$, that is the central
density distribution is
\be\label{eq-distribution}
n(\rho_c) \approx \frac{N_{tot}}{8 \, \rho_0}
\ee
for $2 \, \rho_0 < \rho_c < 10 \, \rho_0$ and 0
otherwise. Here we would like to stress that this simple
choice is not crucial for our conclusions. Indeed, we could 
consider another distribution and get roughly the same result
if the probability of finding a star with a given central 
density is the same within an order of magnitude: for example, 
Eq.~(\ref{eq-distribution}) could be seen as the approximation 
of a gaussian of height $\sim N_{tot}/8 \, \rho_0$ and width 
$\sim 8 \, \rho_0$. Following our first working hypothesis, 
$2 \, \rho_0 < \rho_{crit} < 10 \, \rho_0$
and the possibility of temporal variation of $\Lambda_{QCD}$ 
and $G_N$ may induce the transformation of neutron stars
into strange or hybrid ones, because $\rho_{crit}$ depends on 
$\Lambda_{QCD}$ and $\rho_c$ on $G_N$. In addition to this, 
since we take the phase transition to be of first order, in 
the conversion the compact object has to release a huge 
amount of energy in a very short time, say 10 s. This energy 
is the difference of binding energy between strange/hybrid star 
and neutron one. All the models agree with a typical energy 
release at the level of $10^{52} - 10^{53}$~erg~\cite{ss}.
This huge amount of energy can be converted, at least in
part, to photons and $e^+e^-$ pairs. Various mechanisms
have been proposed~\cite{conversion}, all of them producing
a gamma emission with an energy easily exceeding 
$10^{50} - 10^{51}$~erg. Such a huge power in $\gamma$
can be at the origin of at least a fraction of the so called
long Gamma Ray Bursts (GRBs)~\cite{grbs-rev}.

{\sc Temporal variation of $\Lambda_{QCD}$ --}
A temporal variation of $\Lambda_{QCD}$ must induce
a change in the value of the critical baryon density 
$\rho_{crit}$. A rough estimate can be obtained as
follows. We start by considering two types of quark 
models, the bag-like~\cite{bag} and the NJL-like 
models~\cite{buballa}. In bag models, the crucial parameter
regulating $\rho_{crit}$ is the pressure of the vacuum $B$,
whose numerical value is independent of the presence of
the quarks. Therefore the value of $B$ is controlled by
the only dimensional parameter present also in pure gauge
QCD, i.e. $\Lambda_{QCD}$. One can therefore write
\be\label{lambda}
\rho_{crit} = a \, \Lambda_{QCD}^4 \, ,
\ee
where $a$ is a dimensionless order one coefficient.
In chiral models, and in particular in the NJL-like ones,
gluons are not present as dynamical degrees of freedom.
It is anyway possible to introduce a quantity whose
physical meaning is similar to the one of $B$ in bag
models (see e.g. Ref.~\cite{buballa}). The numerical
value of that quantity is dominated by the value of the
constituent masses of the quarks. In NJL-like models
the mass of the strange quark is so large that it
plays a minor role compared to the up and down quarks.
Since the constituent masses of the $u$ and $d$ quarks
are almost independent of the small value of the current 
mass, one can conclude that also in chiral models the
``pressure of the vacuum'' depends almost only on $\Lambda_{QCD}$.

In the bag model, the relation between $\rho_{crit}$ and $B$ 
has been evaluated in a large number of calculations and is 
typically of the order of~\cite{glende, akmal}
\be
\Delta \rho_{crit} \approx 
\frac{0.003}{\rm MeV} \, \Delta B \, ,
\ee 
where $\Delta \rho_{crit}$ and $\Delta B$ are 
respectively a variation of $\rho_{crit}$ and $B$. From 
Eq.~(\ref{lambda}) we find that 
$\Delta B/B = 4 \, \Delta \Lambda_{QCD}/\Lambda_{QCD}$
and hence
\be
\Delta \rho_{crit} \sim 10 \, \rho_0 \,
\Delta \Lambda_{QCD}/\Lambda_{QCD} \, .
\ee
Let us note that $\Delta \Lambda_{QCD} < 0$ implies 
$\Delta \rho_{crit} < 0$: a time decreasing
$\Lambda_{QCD}$ makes $\rho_{crit}$ decrease as well
and neutron stars with a central density just below 
the critical value may convert into strange or hybrid ones.
The rate of strange/hybrid star production due to the 
possible temporal evolution of $\Lambda_{QCD}$ we can
observe is
\be\label{ss-rate}
\dot{N}_{ss} \sim 
- \epsilon \, n(\rho_c) \, \dot{\rho}_{crit} 
\sim - 10^{20} \, \epsilon \,
\frac{\dot{\Lambda}_{QCD}}{\Lambda_{QCD}} \, .
\ee
Here $\epsilon$ is the fraction of compact stars in the
visible universe whose conversion from hadron to strange
matter can be seen from the Earth and the equation is
valid only for $\dot{\rho}_c < 0$ (hence $\dot{N}_{ss} > 0$).
If $\dot{\rho}_c > 0$, strange stars should convert into
neutron ones, but in this case we do not expect any
violent phenomenon: more probably the picture would 
look like a slow evaporation of strange matter into
hadrons. Because of this, no bound can be deduced
for the case $\dot{\Lambda}_{QCD} > 0$.

In conclusion, according to our theoretical framework,
a time decreasing $\Lambda_{QCD}$ must be accompanied
by spectacular events, that is the formation of
strange or hybrid stars from neutron star progenitors.
As already mentioned, the only known events which can be 
associated to such violent phenomena are the 
GRBs~\cite{grbs-rev}. In particular we are interested
in the so-called long GRBs, which are known to be associated
with the gravitational collapse of massive stars.
Here we would like to stress that 
we are not assuming that all or some of the long GRBs are related to 
the formation of a strange star, but we are saying that the
conversion from a neutron to a strange star, if it occurs 
in the universe, is certainly a quite particular event 
which likely releases a huge amount of energy ($\sim 10^{52}$ erg)
in a short time and that at present we have not yet 
observed other phenomena compatible with this picture
but long GRBs. From this point of view, we can assert that the rate of 
strange star production cannot exceed the one of GRBs. Since 
today the observed GRBs have a mean redshift $z \approx 3$ 
and are more or less uniformly distributed up to redshift 
$z \approx 5$, we would be able to observe most of the 
events in the whole visible universe associated with the 
transition from a neutron star to a strange star. 
This means that we can put $\epsilon \sim 1$ in 
Eq.~(\ref{ss-rate}). In the end, since the observed rate 
of GRBs is
\be
\dot{N}_{GRBs} \approx 10^3 \, {\rm yr}^{-1} \, ,
\ee 
we deduce the following constraint on a possible linearly
time decreasing QCD scale \footnote{Here, as well as in 
Eq.~(\ref{bound-g}) below, the temporal variation is respect
to the cosmological time, that is the time coordinate of the
metric describing the expanding universe. It must be so
because the prediction of time varying ``fundamental constants''
arises when we consider cosmological solutions of theories
beyond general relativity.}
\be\label{bound-qcd}
- \dot{\Lambda}_{QCD}/\Lambda_{QCD}
\lesssim 10^{-17} \, {\rm yr}^{-1} \, .
\ee
One caveat is in order. If the transition from a neutron
star to a hybrid or a quark star produces a beamed emission, the
actual limit on $\dot{N}_{GRBs}$ is larger. GRBs are known
to be beamed, and the correction to the rate due to beaming
is of the order of $75 \pm 25$ \cite{guetta}. If the same beaming
factor applies to GRBs associated with quark deconfinement, than
$\dot{N}_{GRBs} \approx 10^5 \, {\rm yr}^{-1}$ and 
$- \dot{\Lambda}_{QCD}/\Lambda_{QCD} \lesssim 10^{-15} \, {\rm yr}^{-1}$.

{\sc Temporal variation of $G_N$ --}
Let us now consider the possibility of a non-zero ${\dot G}_N$.
Now the critical density $\rho_{crit}$ is unchanged, because 
it is set by QCD physics. However, like all the other stars, 
neutron stars are self-gravitating system whose configuration 
is determined by the balance between the attractive gravitational 
force and the particle pressure. Hence a 
variation of $G_N$ induces a change of the central stellar 
density. Of course, if a neutron star has a central density
just below the critical value, an increase of $G_N$ makes $\rho_c$
pass the critical value and the neutron star can convert into 
a strange one. A rough estimate for the bound on $\dot{G}_N/G_N$
can be deduced from the following simple considerations.
As stellar model, we can take a system of $N$ non-relativistic
fermions of mass $m$ which interact only gravitationally. 
The energy per unit volume is 
\be\label{energy}
\varepsilon \sim 
\frac{3}{5} \, \frac{p_F^2}{2m} \, \rho
+ m \, \rho 
- \frac{3}{5} \frac{G_N m^2 \rho^2 V}{R} \, ,
\ee
where $\rho$ is the particle number density (for the sake of 
simplicity we assume it is constant in the star), 
$p_F = (3 \pi^2 \rho)^{1/3}$ is the Fermi momentum, $R$ the 
stellar radius and $V$ the stellar volume. The first term on 
the right hand side of Eq.~(\ref{energy}) is the particle kinetic 
energy per unit volume, the second term is the particle rest 
energy per unit volume and the last one is the gravitational 
potential energy per unit volume. If we multiply Eq.~(\ref{energy}) by 
$1/\rho$ and then we replace $\rho$ with $N/V$, we get the 
energy per particle as a function of the stellar radius $R$, since
$N$ must be constant. It is straightforward to find that the equilibrium 
radius is proportional to $1/G_N$. In the Newtonian framework, for a 
star with constant density we have $\rho \propto R^{-3} \propto G_N^3$.
At first approximation, we can expect the same dependence
of the central density of compact stars on the gravitational
constant $G_N$ and the relation between a small change of
$G_N$ and $\rho_c$ is
\be
\Delta \rho_c/\rho_c \approx 3 \, \Delta G_N/G_N \, .
\ee
Hence, replacing $\dot{\rho}_{crit}$ with $-\dot{\rho}_c$ 
in Eq.~(\ref{ss-rate}), we find the bound on the temporal 
evolution of $G_N$
\be\label{bound-g}
\dot{G}_N/G_N \lesssim 10^{-17} \, {\rm yr}^{-1} \, .
\ee
As we have already said for the case of the QCD scale 
$\Lambda_{QCD}$, such a bound is applicable only in one
direction, here for an increasing gravitational constant.
However, Eq.~(\ref{bound-g}) is about 5 order of magnitude 
more stringent than the present best constraint of 
Eq.~(\ref{bound-llr}). Also in this case, if the gamma emission
is beamed the limit becomes less stringent, 
$\dot{G}_N/G_N \lesssim 10^{-15} \, {\rm yr}^{-1}$.

{\sc Conclusion --}
In this letter we have discussed the possibility that the
QCD scale $\Lambda_{QCD}$ is decreasing in time or that
the Newton gravitational constant $G_N$ is increasing,
considering their effects on the equilibrium configuration
of compact stars. Under a set of reasonable assumptions,
we have found the constraints
\be\label{conclusion1}
- \dot{\Lambda}_{QCD}/\Lambda_{QCD}
&\lesssim& 10^{-17} \, {\rm yr}^{-1} \, , \\
\dot{G}_N/G_N &\lesssim& 10^{-17} \, {\rm yr}^{-1} \, ,
\label{conclusion2}
\ee
for the case of linearly time varying $\Lambda_{QCD}$ and $G_N$.
Since these bounds come from the non-observation of too much
GRBs, they cover the red-shift range $z \approx 0 - 5$,
that is a time interval of about $10^{10}$ years.

The possibility of time varying
``fundamental constants'' is not exotic, but instead
quite a general prediction of most theories of gravity
beyond general relativity. Assuming that compact stars
can be neutron or strange/hybrid stars, depending on the
value of their central density, we have to expect that
$\dot{\Lambda}_{QCD} < 0$ or $\dot{G}_N > 0$ induces
the transformation of some neutron stars into strange
or hybrid ones. Then, if the phase transition from
hadron to quark matter at such high density is of first
order, as it is commonly believed, the energy release 
is huge and we should be able to
observe the event even if it is at cosmological distant
far from us. At present, the only phenomena compatible
with this picture are the GRBs and, from their rate, we
arrive at the constraints on the temporal evolution of
$\Lambda_{QCD}$ and $G_N$. On the other hand, if the
origin of GRBs were completely different, it would mean that
at present we do not know any phenomenon which can be 
associated to the formation of a strange/hybrid star
from a neutron star and our bounds would become much
stronger. Of course, it is important that future 
investigations will confirm our working hypothesis. 
In the end, we would like to stress that
our constraints (\ref{bound-qcd}) and (\ref{bound-g})
(or (\ref{conclusion1}) and (\ref{conclusion2})) 
are very competitive, even from the point of view of
the covered time interval (at present, only BBN and
CMBR arguments can probe earlier events in the history
of the universe). In particular for the case of the 
gravitational constant, our bound on a linear time 
increasing $G_N$ is much tighter than all the previous 
bounds, even if a possible beaming of the GRB associated
with quark deconfinement is taken into account.
On top of that, the previous bounds 
will be hardly significantly improved in a 
near future for the coming of new measurements and observations. 
On the contrary, our constraint could become much stronger
once the origin of GRBs will be better known.

Finally, it is tempting to discuss a recent analysis indicating
an excess of long GRBs at large red-shift~\cite{kistler}.
More precisely, an evidence has been found of 
a GRB rate  4 times larger than the one predicted assuming the GRB rate to 
follow the star formation rate. A possible interpretation of this 
result can be based on a non-linear evolution of the fundamental constants,
with a more rapid variation in the young universe.

\begin{acknowledgments}
We wish to thank Alexey Petrov for useful comments.
C.B. is supported in part by NSF under grant PHY-0547794 
and by DOE under contract DE-FG02-96ER41005.
\end{acknowledgments}

\end{document}